\documentclass[11pt]{article}

\usepackage[left=2cm, right=2cm, top=2cm, bottom=3cm]{geometry}
\usepackage[utf8]{inputenc}
\usepackage{graphicx}
\usepackage{amsmath}
\usepackage{amssymb}
\usepackage{subcaption}
\usepackage[format=plain]{caption}
\graphicspath{ {./figures}/}
\usepackage{siunitx,physics}
\usepackage[hyphens]{url}
\usepackage{hyperref}
\usepackage[hyphenbreaks]{breakurl}
\usepackage{microtype}
\usepackage[backend=biber,maxnames=10]{biblatex}
\usepackage{xurl}  
\usepackage{authblk} 
\usepackage{booktabs} 
\usepackage{cleveref} 

\newcommand{\Turkey}{T\"urkiye}
\DeclareUnicodeCharacter{2212}{-}

\title{\bf Status of the Proton EDM Experiment (pEDM)}

\author[7]{Jim Alexander}
\author[43]{Vassilis Anastassopoulos}
\author[4]{Grigor Atoian}
\author[33]{Rick Baartman}
\author[46]{Stefan Baeßler}
\author[23]{Franco Bedeschi}
\author[4]{John Benante}
\author[21]{Martin Berz}
\author[4]{Michael Blaskiewicz}
\author[38]{Themis Bowcock}
\author[4]{Kevin Brown}
\author[17,10,9,35]{Dmitry Budker}
\author[38]{Sergey Burdin}
\author[8]{Brendan C. Casey}
\author[40]{Gianluigi Casse}
\author[45]{Giovanni Cantatore}
\author[40]{Timothy Chupp}
\author[4]{Hooman Davoudiasl}
\author[4]{Dmitri Denisov}
\author[4]{Bhawin Dhital}
\author[4]{Milind V. Diwan}
\author[37]{Renee Fatemi}
\author[24]{George Fanourakis}
\author[4]{Wolfram Fischer}
\author[31]{Peter Graham}
\author[28]{Frederick Gray}
\author[43]{Antonios Gardikiotis}
\author[22]{Claudio Gatti}
\author[38]{James Gooding}
\author[13,36]{Boxing Gou}
\author[14]{Selcuk Haciomeroglu}
\author[7]{Georg H. Hoffstaetter}
\author[4]{Haixin Huang}
\author[23]{Marco Incagli}
\author[27]{Hoyong Jeong}
\author[18]{David Kaplan}
\author[44]{Marin Karuza}
\author[34]{David Kawall}
\author[39]{Alexander Keshavarzi}
\author[41]{On Kim}
\author[17,10]{Younggeun Kim}
\author[5]{Ivan Koop}
\author[19]{Valeri Lebedev}
\author[32]{Jonathan Lee}
\author[6]{Soohyung Lee}
\author[23,30]{Alberto Lusiani}
\author[4]{William J. Marciano}
\author[12]{Marios Maroudas}
\author[20]{Andrei Matlashov}
\author[4]{Francois Meot}
\author[3]{James P. Miller}
\author[4]{William M. Morse}
\author[8]{James Mott}
\author[26]{Zhanibek Omarov}
\author[15]{Cenap Ozben}
\author[42]{Giovanni Maria Piacentino}
\author[16]{Matthew Poelker}
\author[46]{Dinko Pocanic}
\author[4]{Boris Podobedov}
\author[38]{Joe Price}
\author[4]{Xin Qian}
\author[18]{Surjeet Rajendran}
\author[4]{Deepak Raparia}
\author[4]{Sergio Rescia}
\author[3]{B. Lee Roberts}
\author[26]{Yannis K. Semertzidis}
\author[19]{Alexander Silenko}
\author[4]{Amarjit Soni}
\author[11]{Edward Stephenson}
\author[16]{Riad Suleiman}
\author[25]{Michael Syphers}
\author[29]{Pia Thoerngren}
\author[4]{Volodya Tishchenko}
\author[4]{Nicholaos Tsoupas}
\author[2]{Spyros Tzamarias}
\author[22]{Alessandro Variola}
\author[23,38]{Graziano Venanzoni}
\author[38]{Eva Vilella}
\author[38]{Joost Vossebeld}
\author[1]{Peter Winter}
\author[16,47]{Bogdan Wojtsekhowski}
\author[27]{Eunil Won}
\author[43]{Konstantin Zioutas}

\affil[1]{%
  Argonne National Laboratory, Lemont, Illinois, USA}
\affil[2]{%
  Aristotle University of Thessaloniki, Thessaloniki, Greece}
\affil[3]{%
  Boston University, Boston, Massachusetts, USA}
\affil[4]{%
  Brookhaven National Laboratory, Upton, New York, USA}
\affil[5]{%
  Budker Institute of Nuclear Physics, Novosibirsk, Russia}
\affil[6]{%
  Center for Accelerator Research, Korea University, Sejong, Republic of Korea}
\affil[7]{%
  Cornell University, Ithaca, New York, USA}
\affil[8]{%
  Fermi National Accelerator Laboratory, Batavia, Illinois, USA}
  \affil[9]{%
  GSI Helmholtzzentrum für Schwerionenforschung GmbH, Darmstadt, Germany}
\affil[10]{%
  Helmholtz Institute Mainz, Mainz, Germany}
\affil[11]{%
  Indiana University, Bloomington, Indiana, USA}
\affil[12]{%
  Institute of Experimental Physics, University of Hamburg, Hamburg, Germany}
\affil[13]{%
  Institute of Modern Physics, Chinese Academy of Sciences, Lanzhou, China}
\affil[14]{%
  Istinye University, Istanbul, \Turkey}
\affil[15]{%
  Istanbul Technical University, Istanbul, \Turkey}
\affil[16]{%
Jefferson Laboratory, Newport News, Virginia, USA}
  \affil[17]{%
  Johannes Gutenberg-Universit{\"a}t Mainz, Mainz, Germany}
\affil[18]{%
  Johns Hopkins University, Baltimore, Maryland, USA}
\affil[19]{%
  Joint Institute for Nuclear Research, Dubna, Russia}
\affil[20]{%
  Los Alamos National Laboratory, Los Alamos, New Mexico, USA}
\affil[21]{%
  Michigan State University, East Lansing, Michigan, USA}
\affil[22]{%
  National Institute for Nuclear Physics (INFN-Frascati), Rome, Italy}
\affil[23]{%
  National Institute for Nuclear Physics (INFN-Pisa), Pisa, Italy}
\affil[24]{%
  NCSR Demokritos Institute of Nuclear and Particle Physics, Athens, Greece}
\affil[25]{%
  Northern Illinois University, DeKalb, Illinois, USA}
\affil[26]{%
  Physics Department, KAIST, Daejeon, Republic of Korea}
\affil[27]{%
  Physics Department, Korea University, Seoul, Republic of Korea}
\affil[28]{%
  Regis University, Denver, Colorado, USA}
\affil[29]{%
  Royal Institute of Technology, Division of Nuclear Physics, Stockholm, Sweden}
\affil[30]{%
  Scuola Normale Superiore di Pisa, Pisa, Italy}
\affil[31]{%
  Stanford University, Stanford, California, USA}
\affil[32]{%
  Stony Brook University, Stony Brook, New York, USA}
\affil[33]{%
  TRIUMF, Vancouver, British Columbia, Canada}
\affil[34]{%
  UMass Amherst, Amherst, Massachusetts, USA}
\affil[35]{%
  University of California, Berkeley, California, USA}
\affil[36]{%
  University of Chinese Academy of Sciences, Beijing, China}
\affil[37]{%
  University of Kentucky, Lexington, Kentucky, USA}
\affil[38]{%
  University of Liverpool, Liverpool, UK}
\affil[39]{%
  University of Manchester, Manchester, UK}
\affil[40]{%
  University of Michigan, Ann Arbor, Michigan, USA}
\affil[41]{%
  University of Mississippi, University, Mississippi, USA}
\affil[42]{%
  University of Molise, Campobasso, Italy}
\affil[43]{%
  University of Patras, Dept. of Physics, Patras-Rio, Greece}
\affil[44]{%
  University of Rijeka, Rijeka, Croatia}
\affil[45]{%
  University of Trieste and National Institute for Nuclear Physics (INFN-Trieste), Trieste, Italy}
\affil[46]{%
  University of Virginia, Charlottesville, Virginia, USA}
\affil[47]{%
William \& Mary College, Williamsburg, Virginia, USA}

\begin{document}

\maketitle

\begin{abstract}

The Proton EDM Experiment (pEDM) is the first direct search for the proton electric dipole moment (EDM) with the aim of being the first experiment to probe the Standard Model (SM) prediction of any particle EDM. Phase-I of pEDM will achieve $10^{-29} e\cdot$cm, improving current indirect limits by four orders of magnitude. This will establish a new standard of precision in nucleon EDM searches and offer a unique sensitivity to better understand the Strong CP problem. The experiment is ideally positioned to explore physics beyond the Standard Model (BSM), with sensitivity to axionic dark matter via the signal of an oscillating proton EDM and across a wide mass range of BSM models from $\mathcal{O}(1\text{GeV})$ to $\mathcal{O}(10^3\text{TeV})$. Utilizing the frozen-spin technique in a highly symmetric storage ring that leverages existing infrastructure at Brookhaven National Laboratory (BNL), pEDM builds upon the technological foundation and experimental expertise of the highly successful Muon $g$$-$$2$ Experiments. With significant R\&D and prototyping already underway, pEDM is preparing a conceptual design report (CDR) to offer a cost-effective, high-impact path to discovering new sources of CP violation and advancing our understanding of fundamental physics. It will play a vital role in complementing the physics goals of the next-generation collider while simultaneously contributing to sustaining particle physics research and training early-career researchers during gaps between major collider operations.
\break

\noindent Document submitted as part of the US National Input to the European Strategy for Particle Physics Update 2026 (\url{https://indico.cern.ch/event/1439855/contributions/6461648/}).

\end{abstract}

\section{Executive summary}  

\begin{figure*}[!t]
  \centering
  \begin{subfigure}[t]{0.41\textwidth}
    \includegraphics[width=\linewidth]{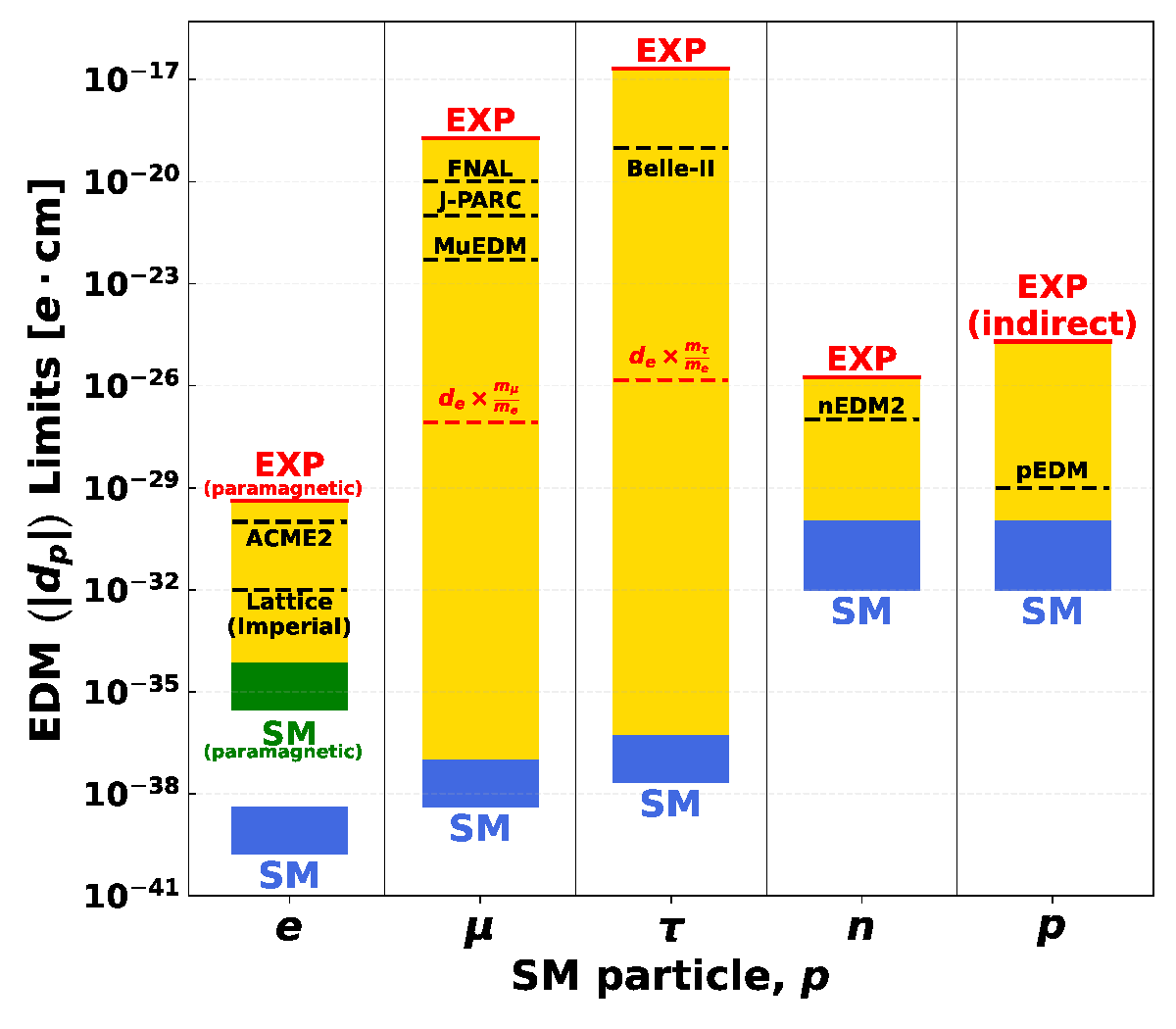}
    \caption*{(a)}
  \end{subfigure}
  \hspace{0.1cm}
  \begin{subfigure}[t]{0.53\textwidth}
    \includegraphics[width=\linewidth]{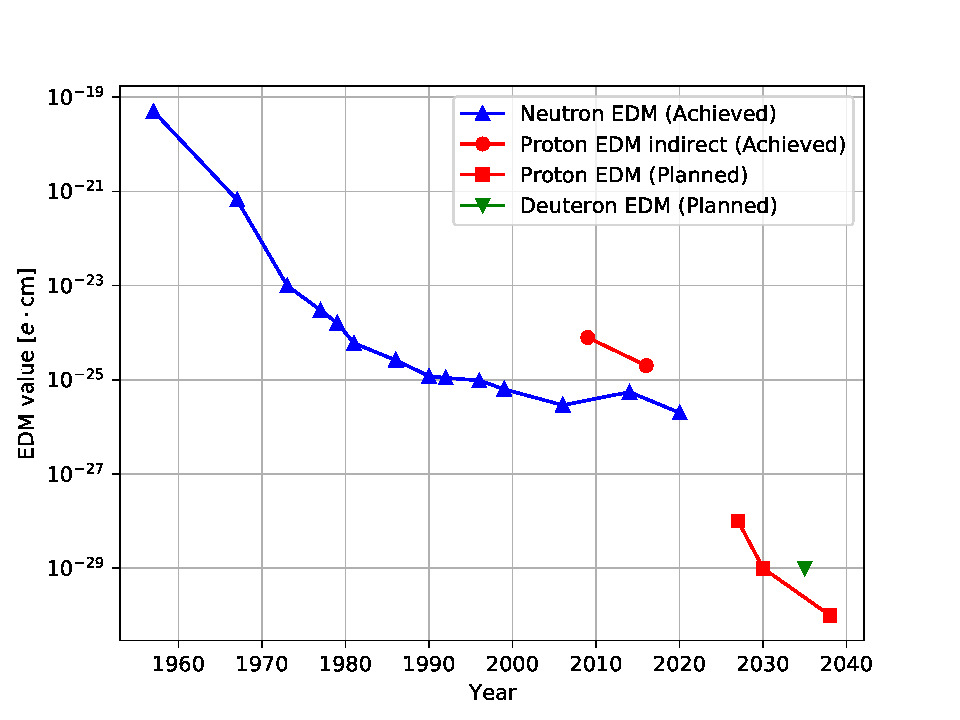}
    \caption*{(b)}
     \vspace{0.3cm}
  \end{subfigure}
  \caption{From left to right: (a) Limits on the EDM of the electron, muon, tau, neutron and proton. The yellow bands show the range of potential BSM interactions, down to the SM limit (blue line). The red lines show each current experimental (EXP) limit. The black dotted lines show the expected sensitivity from proposed experiments. (b) The neutron and proton (indirect) EDM limits, and pEDM projected sensitivities for the proton and  deuteron nuclei, shown as a function of publication year. The planned target dates reflect a technically driven schedule.}
  \label{fig:limits}
\end{figure*}

The Proton EDM Experiment (pEDM) is the first direct search for the proton electric dipole moment (EDM)~\cite{Alexander:2022rmq}, $d_p$, with the aim of being the first experiment to probe the Standard Model (SM) prediction of any particle EDM. The SM prediction of the proton EDM is $d_p^{\rm SM} \sim10^{-31} e\cdot$cm~\cite{CPEDM:2019nwp}. The intended phase-I of pEDM will achieve a new paradigm of precision in nucleon EDMs by reaching $10^{-29} e\cdot$cm, improving the current indirect limit by $\mathcal{O}(10^4)$~\cite{Alexander:2022rmq}. As shown in Fig.~\ref{fig:limits}, it will cover much of the unexplored range for new physics and far surpass all other EDM measurement efforts in probing down to the SM prediction for an EDM. pEDM is one of the best hopes to discover a new source of CP violation to help explain the universe's baryon asymmetry and to address the Strong CP problem by improving on the current sensitivity to the QCD $\theta$-term by $\mathcal{O}(10^3)$ in phase-I. 
pEDM has a BSM sensitivity ranging from $\mathcal{O}(1 {\rm ~GeV}) \rightarrow \mathcal{O}(10^3 {\rm ~TeV})$, making it complementary to light, weak new physics searches at e.g. LZ~\cite{LZ:2019sgr}, LDMX~\cite{LDMX:2018cma}, FASER~\cite{Feng:2017uoz} and SHiP~\cite{SHiP:2015vad} and energy-frontier searches for heavy new physics at the LHC~\cite{Butterworth:2021jto,ATLAS:2008xda,CMS:2008xjf,LHCb:2008vvz,ALICE:2008ngc}, and proposed FCC~\cite{FCC:2018byv} and muon collider programs~\cite{Black:2022cth}. Crucially, an oscillating proton EDM signal would be evidence for axionic dark matter (DM)~\cite{Chadha-Day:2021szb,Semertzidis:2021rxs,Co:2020xlh,DiLuzio:2021gos,Schulthess:2022pbp}, with pEDM phase-I alone being sensitive to a wide range of axion frequencies $(1 {\rm mHz} \rightarrow 1 {\rm MHz})$ and masses $(10^{-18}{\rm eV} \rightarrow 10^{-9}{\rm eV})$~\cite{Alexander:2022rmq}.

pEDM is a charged particle storage-ring experiment which will employ the novel frozen-spin technique~\cite{Adelmann:2021udj,Alexander:2022rmq} to precisely measure the vertical rotation of the polarization of a highly-polarized stored proton beam. pEDM's experimental techniques are based upon the highly successful Muon $g$$-$$2$ Experiment at Fermilab~\cite{Muong-2:2023cdq,Muong-2:2024hpx,Muong-2:2021ojo,Muong-2:2021xzz,Muong-2:2021ovs,Muong-2:2021vma} (and predecessor Muon $g$$-$$2$ Experiments at Brookhaven National Laboratory (BNL)~\cite{Muong-2:2006rrc,Muong-2:2002wip,Muong-2:2004fok} and CERN~\cite{CERN-Mainz-Daresbury:1978ccd,Combley:1974tw}), which in many ways can be considered as a prototype experiment for pEDM. Additionally, significant prototyping of pEDM was completed at the COSY ring in Jülich (Germany)~\cite{CPEDM:2019nwp} and is ongoing at BNL and in the UK. Consequently, pEDM phase-I is in an advanced stage of development for construction of a highly-symmetric proton storage ring at BNL within the AGS tunnel. 

Significant studies concerning the viability of pEDM have been performed over several decades and can be found in e.g.~\cite{Zurek:2019yhj,Morse:2013hoa,Metodiev:2014ola,BASE:2014drs,JEDI:2015vwa,Metodiev:2015twa,JEDI:2016swi,Anastassopoulos:2015ura,JEDI:2017wlr,JEDI:2017bnp,JEDI:2018txs,JEDI:2017lbv,Haciomeroglu:2018nre,Omarov:2020kws,Rathmann:2013rqa,Gooding:2022aks}. Importantly, prototypes of the pEDM's storage ring vacuum chambers, electric field bending, and proton polarimeters are currently being constructed and tested, and a results-based conceptual design report (CDR) is in preparation to be published by the end of the 2026. The ring and its optics will take 3--5 years of construction, 2--3 years statistics collection are needed to reach the first physics publication, and a total of 5 years phase-I statistics collection will reach $10^{-29}\, e\cdot$cm without using stochastic cooling (SC) and less than half a year with SC (currently under evaluation). pEDM can run symbiotically with both the EIC and AGS operations program at BNL, and there is significant development overlap with the EIC and its physics goals to be exploited e.g. in proton polarimetry. The construction cost of pEDM phase-I is estimated to be $\mathcal{O}(\$100{\rm M})$. R\&D for phase-II is underway with the specific aim to utilize the phase-I infrastructure to be sensitive to the SM prediction of $\sim 10^{-31}\, e\cdot$cm. In this scenario, pEDM would therefore either discover or rule out the existence of an EDM with a magnitude above the SM’s highly suppressed value for the first time. 

In the context of the upcoming 2026 European Strategy for Particle Physics Update (ESPPU)~\cite{EuropeanStrategyGroup:2020pow}, pEDM is a prime candidate to complement Europe’s physics beyond colliders (PBC) strategy~\cite{Alemany:2019vsk} to extend discoveries at energy-frontier colliders. In a world where pushing the boundaries of particle physics discovery means weighing the benefits of scientific progress with both financial and environmental cost, pEDM will leverage existing accelerator and experimental infrastructure at BNL to offer a cost-effective yet powerful approach to expand the current and future physics landscape, enhancing sensitivity to crucial new physics that is complementary to the future of energy-frontier colliders. Like the Muon $g$$-$$2$ Experiment at Fermilab, pEDM phase-I is expected to go from TDR to final publication in $<$ 20 years. Smaller, shorter-timescale experiments like pEDM are particularly important when considering the future for early-career researchers (ECRs) in particle physics, who face the prospect of potentially longer experimental timescales than particle physics has ever experienced, and e.g. gaps in CERN's collider program between the end of the HL-LHC and start of the FCC. 

pEDM offers a viable and attractive career path in order to sustain and develop the particle physics population and expertise, and provide continuity in parallel to and in-between major collider projects in training ECRs to develop critical skills, ensuring that we remain a community of researchers with great diversity of experimental and theoretical expertise.

\section{Experimental technique}

 The proposed method has its origins in the measurements of the anomalous magnetic moment of the muon ~\cite{Muong-2:2023cdq,Muong-2:2024hpx,Muong-2:2021ojo,Muong-2:2021xzz,Muong-2:2021ovs,Muong-2:2021vma,Muong-2:2006rrc,Muong-2:2002wip,Muong-2:2004fok,CERN-Mainz-Daresbury:1978ccd,Combley:1974tw}, where an electric field at the so-called ``magic'' momentum does not influence the particle $(g-2)$ precession. Rather, the electric field precesses the momentum and the spin at exactly the same rate, so the difference is zero. The fact that all electric fields have this feature opened up  the possibility of using electric quadrupoles in the ring to focus the beam, while the magnetic field is kept uniform. 
 
 The precession rate of the longitudinal component of the spin in a storage ring with electric and magnetic fields is given by:
\begin{equation}
  \dv{\vec{\beta} \cdot \vec{s}}{t} =
  -\frac{e}{m} \vec{s}_{\perp} \cdot \left[
    \left( \frac{g - 2}{2} \right) \hat{\beta} \times \vec{B}
    + \left( \frac{g \beta}{2} - \frac{1}{\beta} \right) \frac{\vec{E}}{c}
  \right]
  \label{eq:omega}
\end{equation}

The storage ring/magic momentum technique gained a factor of \num{2e3} in systematic error for the muon $g$$-$$2$ experiments. Using these techniques, the BNL muon $g$$-$$2$ experiment set a limit on the EDM of the muon: $d_{\mu} < \num{1.9e-19}~ e \cdot \textnormal{cm}$~\cite{bnl_edm}. For FNAL muon $g$$-$$2$ experiment, we expect this result to improve by up to two orders of magnitude. The statistical and systematic errors for the muon EDM will then be roughly equal. The dominant systematic error effect is due to radial magnetic fields. 

\begin{figure}[!t]
    \centering
    \begin{subfigure}[t]{0.57\textwidth}
        \includegraphics[width=\linewidth]{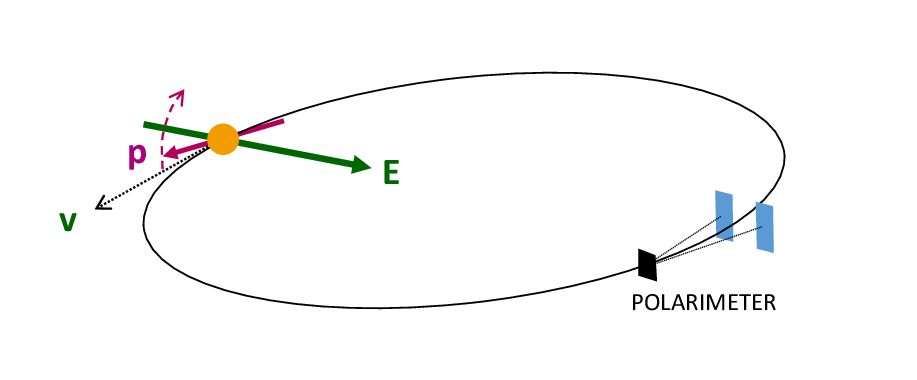} 
    \end{subfigure}
    \hspace{-1cm} 
    \begin{subfigure}[t]{0.43\textwidth}
        \includegraphics[width=\linewidth]{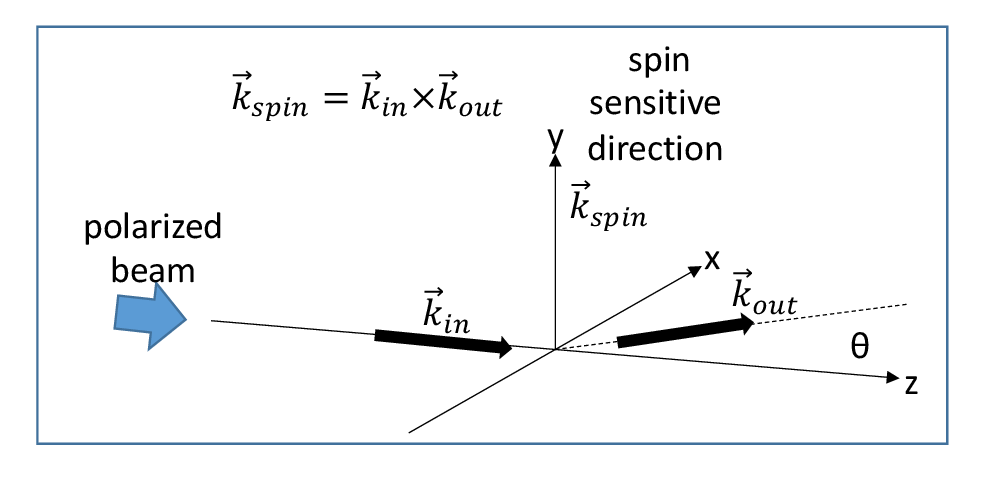} 
    \end{subfigure}
        \caption{The pEDM storage ring concept, with the horizontal spin precession locked to the momentum precession rate (frozen-spin). The radial electric field acts on the particle EDM for the duration of the storage time.  Positive and negative helicity bunches are stored, as well as bunches with their polarization pointing in the radial direction, for systematic error cancellations. In addition, simultaneous clockwise and counterclockwise storage is used to cancel the main systematic errors. The ring circumference is $\sim\SI{800}{m}$. The right inset shows the cross section geometry that is enhanced in parity-conserving Coulomb and nuclear scattering as the EDM signal increases over time.}
    \label{fig:EJS_ring}
\end{figure}

For pEDM, a storage ring at the proton magic momentum (see Table~\ref{tab:magicgamma}) with electric bending and magnetic focusing will be used. This gives a negligible radial magnetic field systematic effect, while the dominant (main) systematic errors effectively cancel by exploiting simultaneous clockwise and counterclockwise storage. The concept of the storage ring EDM experiment is illustrated in \Cref{fig:EJS_ring}. There are three starting requirements: (1) The proton beam must be highly polarized in the ring plane. (2) The frozen-spin technique requires that the momentum of the beam must match the magic value of $p=0.7007$~GeV/c, where the ring-plane spin precession is the same as the velocity precession. (3) The polarization must initially be aligned with the beam velocity axis and maintained along that direction through feedback during storage.

\begin{table}[!b]
  \centering
  \begin{tabular}{lllll}
    \toprule
    $G$           & $\beta$         & $\gamma$        & $p$                  & $KE$
    \\
    \midrule
    $\num{1.793}\qq{}$ & $\num{0.598}\qq{}$ & $\num{1.248}\qq{}$ & $\SI{0.7}{GeV/c}\qq{}$ & $\SI{233}{MeV}$\\
    \bottomrule
  \end{tabular}
    \caption{Magic momentum parameters for protons~\cite{mooser_direct_2014}.}
  \label{tab:magicgamma}
    \vspace{-0.3cm}
\end{table}

The electric field acts along the radial direction toward the center of the ring (E). It is perpendicular to the spin axis (p) and therefore perpendicular to the axis of the EDM. In this situation, the spin will precess in the vertical plane as shown in \Cref{fig:EJS_ring}. The appearance of a vertical polarization component with time is the signal for a non-vanishing EDM. This signal is measured at the polarimeter where a sample of the beam is continuously brought to a carbon target. Elastic proton scattering is measured by two downstream detectors (shown in blue)~\cite{hom88,brantjes2012correcting}. The rates depend on the polarization component $p_y$ because it is connected to the {\textit{axial}} vector created from the proton momenta $\vec{k}_{in}\times\vec{k}_{out}$. The sign of $p_y$ flips between left and right as it follows the changing direction of $\vec{k}_{out}$. Thus, the asymmetry in the left-right rates, $(L-R)/(L+R)=p_yA$, is proportional to $p_y$ and hence the magnitude of the EDM. Here $A$ is the scattering-angle-dependent \emph{analyzing power}, a property of the scattering process. Having both left and right rates together reduces systematic errors.

The pEDM collaboration is also investigating a longer-term possibility of measuring the EDM of the antiproton in the same storage ring, in addition to deuteron and $^3$He nuclei. If this becomes possible (when the pEDM ring is collocated with a source of polarized antiprotons), it will both provide an additional ``reversal'' for the control of systematics and will enable precise comparison of the proton and antiproton EDMs as a unique test of CPT and a probe for ``asymmetric’’ dark matter scenarios~\cite{Budker25}. 

\section{The hybrid-symmetric storage ring}

The selected storage ring design for this experiment is the hybrid-symmetric configuration, which has been studied extensively, and has achieved $10^{-29}~ e \cdot \textnormal{cm}$  sensitivity, all using currently available technologies. Details of this design are described below and can be found in~\cite{Haciomeroglu:2018nre,Omarov:2020kws}.

Based on an all-electric ring concept, the design replaces electric focusing with alternating-gradient magnetic focusing, while maintaining simultaneous CW and CCW beam storage. A major advantage of this design is its enhanced ring-lattice symmetry, which inherently suppresses one of the leading systematic error sources: the average vertical beam velocity in the bending sections~\cite{Omarov:2020kws}. Moreover, alternating magnetic focusing enables strong vertical focusing, improving lattice phase-space acceptance, extending the intrabeam scattering (IBS) lifetime, and reducing sensitivity to external magnetic fields. The key features of the hybrid-symmetric, frozen-spin lattice are summarized in Table~\ref{tab:hybridsymmetric}.

\begin{table*}[!t]
\small
  \centering
 
\begin{tabular}[t]{p{0.45\linewidth} p{0.45\linewidth}}
 \toprule
 Feature  & Comment \\ \midrule
 CW and CCW beam storage simultaneously. 
 & Eliminates the primary systematic error source: the vertical electric field (when magnetic focusing is used). \vspace{0.1cm}\\
 
 Balanced currents in CW and CCW beam. 
 & To optimize systematic error suppression, current cancellation must reach the $10^{-4}$ level or better per injection. The total DC current should be measured to be zero within this level of uncertainty or better.\vspace{0.1cm}\\ 
 
 Longitudinally polarized beams with both helicities. 
 & Suppresses polarimeter related systematic errors, which are primarily caused by beam motion on the polarimeter target.\vspace{0.1cm}\\
 
 Radially polarized beams with both polarization directions. 
 & Enables probing of vector dark matter/energy and is $>10^3$ times more sensitive to the vertical velocity systematic error source than longitudinally polarized beams.\vspace{0.1cm}\\
 
 Current flip of focusing (magnetic) quadrupoles. 
 & Key control parameter for mitigating the vertical velocity effect arising from localization uncertainties in the magnetic quadrupoles. \vspace{0.1cm}\\
 
Spin-based alignment~\cite{Omarov:2020kws}. & The dominant systematic error results from the interplay between unintended, design-limited vertical electric focusing in the bending plates and corresponding harmonics of the radial magnetic field background. These effects are probed and suppressed one harmonic at a time, up to harmonic 24, with the lower-order harmonics contributing most significantly.
 \vspace{0.1cm}\\
 
 Precision control of the beam location. & Beam planarity must be maintained to \SI{0.1}{mm}, with average beam splitting between the counter-rotating (CR) beams kept below \SI{0.01}{mm} on average throughout the experiment runs. \\
 
\bottomrule
\end{tabular}
  \caption{Key Features of the hybrid-symmetric, frozen-spin storage ring design}\label{tab:hybridsymmetric} 
      \vspace{-0.3cm}
\end{table*}

The hybrid-symmetric, frozen-spin method successfully addresses all known sources of potential systematic errors. It effectively eliminates the dominant systematic error—the vertical electric field—while also suppressing the influence of the radial magnetic field to a manageable level. This significantly reduces, and potentially eliminates, the need for magnetic shielding of the ring from external disturbances. Hosting the ring in an existing large tunnel (e.g. the AGS at BNL) further lowers the required radial electric field strength to below 4.5~MV/m, a level expected to be achieved with a cold cathode (dark) current of less than $1\,\mu$A, and possibly even below $1\,$nA, thereby removing the need for high-precision dark current monitoring.

Due to the use of alternating-gradient focusing, strong vertical focusing is achievable, making beam storage robust and enabling an intrabeam scattering (IBS) lifetime of hundred seconds, even without SC. The spin coherence time (SCT) is estimated to exceed $10^3$ seconds, and both parameters are expected to increase significantly with SC, potentially reaching effectively ``infinite'' durations, which would enable excellent statistical sensitivity well beyond $10^{-29}\,e\cdot$cm.

Once the ring is constructed, it provides a unique opportunity to experimentally evaluate whether systematic errors can be brought further under control using different configurations of magnetic and electric fields than those proposed for the hybrid-symmetric, frozen-spin. If so, the ring can be reconfigured to potentially improve the sensitivity of the proton EDM by more than an order of magnitude and bringing $\sim10^{-30}\,e\cdot$cm range within reach even without SC.

A highly symmetric lattice is necessary to limit the EDM and dark matter/dark energy systematics, see~\cite{graham_paper,Omarov:2020kws}. The 24-fold symmetric ring parameters are given in \Cref{tab:specs}. The ring circumference is approximately \SI{800}{m}, with bending electric field \SI{4.4}\,{\rm MV/m}. This circumference is that of the BNL AGS tunnel, the use of which would save tunnel construction costs. $E=\SI{4.4}\,{\rm MV/m}$ is conservative. A pEDM experiment at another location could have up to $E=\SI{5}\,{\rm MV/m}$ without requiring additional R\&D, see~\cite{electrode1,electrode2,electrode3}, thereby setting the scale for the required ring circumference.

\begin{table}[!t]
  \centering
  \begin{minipage}[t]{0.45\textwidth}
    \centering
    \begin{tabular}[t]{lc}
      \toprule
      Quantity & Value \\
      \midrule
      Bending Radius $R_{0}$ & \SI{95.49}{m} \\
      Number of periods & 24 \\
      Electrode spacing & \SI{4}{cm} \\
      Electrode height & \SI{20}{cm} \\
      Deflector shape & cylindrical \\
      Radial bending $E$-field & \SI{4.4}{MV/m} \\
      Straight section length & \SI{4.16}{m} \\
      Quadrupole length & \SI{0.4}{m} \\
      Quadrupole strength & \SI{\pm 0.21}{T/m} \\
      Bending section length & \SI{12.5}{m} \\
      Bending circumference & \SI{600}{m} \\
      Total circumference & \SI{800}{m} \\
      Cyclotron frequency & \SI{224}{kHz} \\
      Revolution time & \SI{4.46}{\micro s} \\
      $\beta_{x}^{\textnormal{max}},~ \beta_{y}^{\textnormal{max}}$ & \SI{64.54}{m}, \SI{77.39}{m} \\
      Dispersion, $D_{x}^{\textnormal{max}}$ & \SI{33.81}{m} \\
      \bottomrule
    \end{tabular}
  \end{minipage}%
  \hfill
  \begin{minipage}[t]{0.55\textwidth}
    \centering
    \begin{tabular}[t]{lc}
      \toprule
      Quantity & Value \\
      \midrule
      Tunes, $Q_{x}, ~ Q_{y}$ & 2.699, 2.245 \\
      Slip factor, $\frac{dt}{t}/\frac{dp}{p}$ & -0.253 \\
      Momentum acceptance, $(dp/p)$ & \num{5.2e-4} \\
      Horizontal acceptance [\si{mm} \si{mrad}] & 4.8 \\
      RMS emittance [\si{mm} \si{mrad}], $\epsilon_{x}, ~\epsilon_{y}$ & 0.214, 0.250 \\
      RMS momentum spread & \num{1.177e-4} \\
      Particles per bunch & \num{1.17e8} \\
      RF voltage & \SI{1.89}{kV} \\
      Harmonic number, $h$ & 80 \\
      Synchrotron tune, $Q_{s}$ & \num{3.81e-3} \\
      Bucket height, $\Delta p/p_{\textnormal{bucket}}$ & \num{3.77e-4} \\
      Bucket length & \SI{10}{m} \\
      RMS bunch length, $\sigma_{s}$ & \SI{0.994}{m} \\
      Beam planarity & \SI{\le 0.1}{mm} \\
      CR-beam splitting & \SI{\le 0.01}{mm} \\
      \vspace{-0.07cm}\\
      \bottomrule
    \end{tabular}
  \end{minipage}
  \caption{Ring and beam parameters for the hybrid-symmetric ring design. The proton beam polarization is 80\%. The beam planarity refers to the vertical orbit of the counter-rotating (CR) beams with respect to gravity around the ring.}
  \label{tab:specs}
  \vspace{-0.1cm}
\end{table}

\section{Status and ongoing efforts}

pEDM's CDR is currently under preparation with completion expected within a year.
Key capability targets include:
\begin{itemize}
  \item High-voltage plate testing to a minimum of 5 MV/m (operational target: $4.4$~MV/m), with cold-cathode current less than \SI{1}{\micro\ampere}.
  \item Beam polarization lifetime (SCT) exceeding $10^3$ s.
  \item High-efficiency spin and beam dynamics simulations with $10^3$ particles.
  \item Construction of a high-purity magnetic quadrupole capable of providing opposite fields at every consecutive injection.
  \item Capability for multiple EDM measurements (proton, deuteron, $^3$He, and possibly even antiproton) within 20 years.
  \item Systematic error reduction, feasibility studies, and minimizing duplication. 
\end{itemize}
The major milestones that will be demonstrated for the CDR include:
\begin{itemize}
  \item Final design of the deflector chamber and high-voltage electric plates (TiN-coated), with cold-cathode current less than \SI{1}{\nano\ampere}.
  \item Continued simulations of spin and beam dynamics with $10^3$ particles for more than 1 million revolutions.
  \item Exploration of advanced techniques like stochastic cooling to extend SCT to more than $10^5$ s.
  \item Engagement with international partners for design validation and subsystem contributions.
  \item Development of an ultra-pure magnetic quadrupole that avoids ferromagnetic cores to prevent hysteresis in the quadrupole magnets, followed by measurement of its harmonic content. 
\end{itemize}

\begin{figure}[!t]
    \centering
    \begin{subfigure}[t]{0.95\textwidth}
        \includegraphics[width=\linewidth]{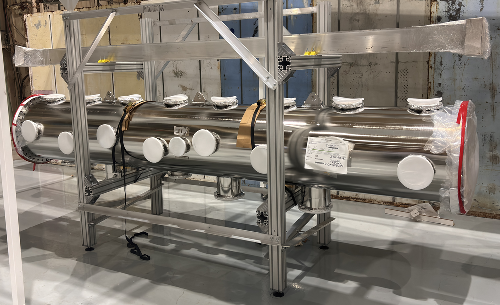} 
        \caption{The prototype storage ring vacuum chamber at BNL in November 2024.}
        \vspace{.2cm}
        \label{fig:chamber}
    \end{subfigure}
    \begin{subfigure}[t]{0.48\textwidth}
        \includegraphics[trim=50 0 50 0, width=\linewidth]{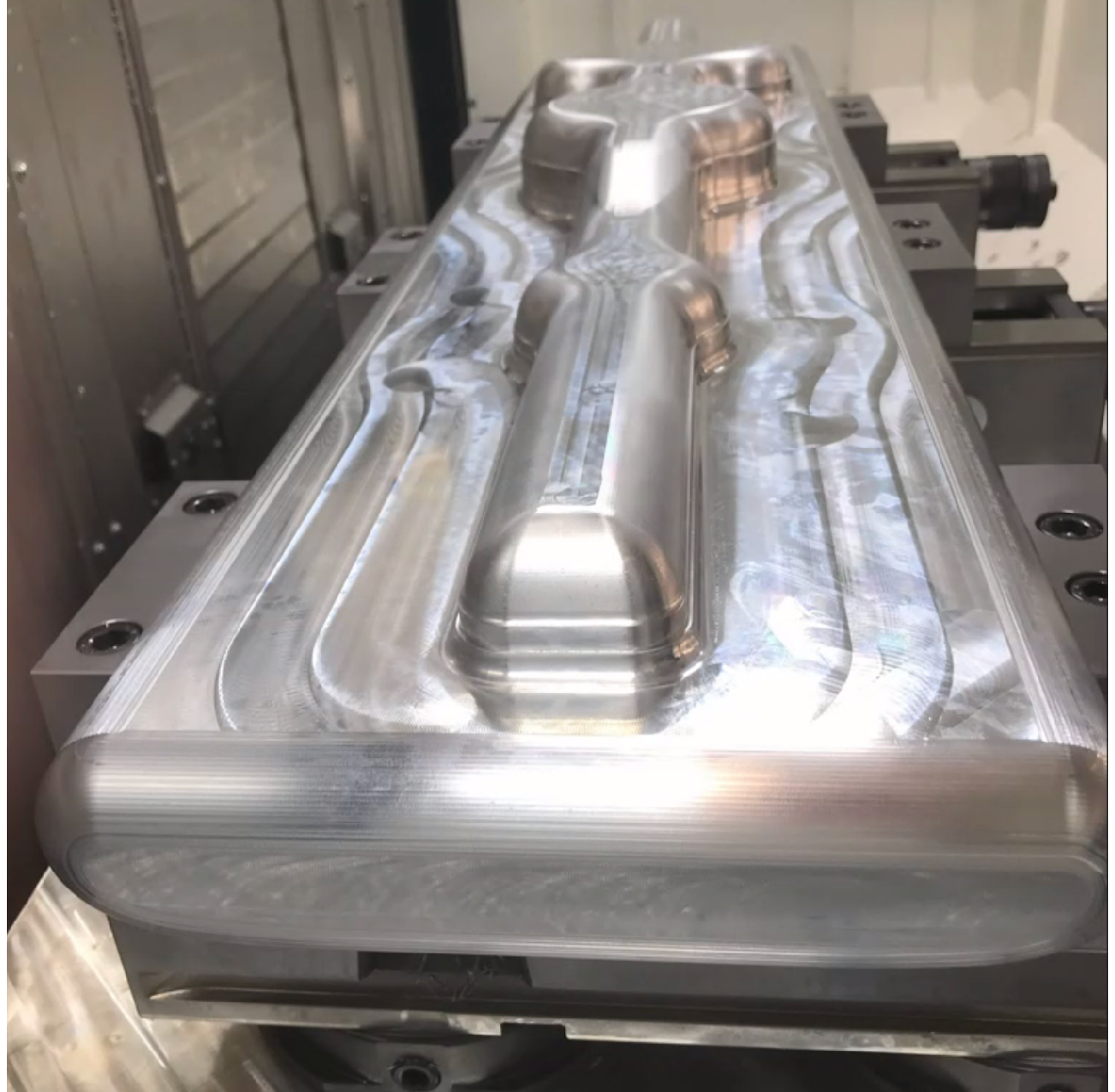} 
        \caption{A precision-machined high-voltage electrode prototype at University of Liverpool (UK) in February 2025.}
        \label{fig:electrode}
    \end{subfigure}
    \hspace{0.2cm}
        \begin{subfigure}[t]{0.48\textwidth}
        \includegraphics[trim=50 0 50 0, width=\linewidth]{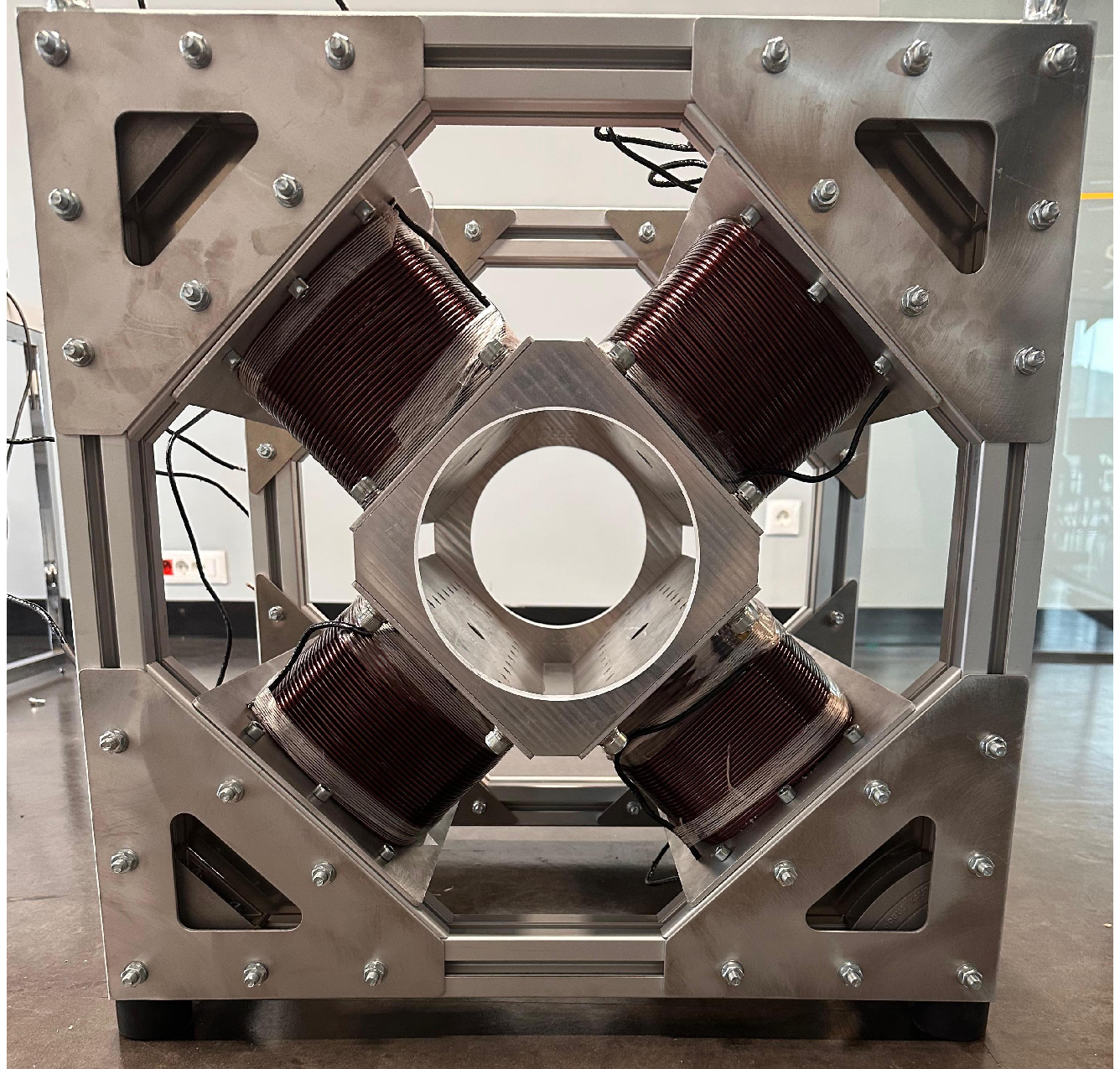} 
        \caption{A full-scale magnetic quadrupole prototype assembly at Istinye University (\Turkey) in March 2025.}
        \label{fig:Mag_quad}
    \end{subfigure}
    \caption{Recent photographs of prototype components of pEDM's hybrid-symmetric storage ring.}
    \label{fig:prototyped}
    \vspace{-0.3cm}
\end{figure}

The USA and UK are collaborating to prototype the deflector chamber and high-voltage electrode plates. A prototype deflector chamber (see Figure~\ref{fig:chamber}) was designed and constructed in the USA in late 2024 and is being prepared for the insertion of the electrodes, operation under vacuum, and high-voltage testing. The electrode plates are being precision-machined in the UK to mechanical and metrological specification (e.g. flat to $<$ \SI{10}{\micro\meter} over \SI{1}{m}) using pneumatic metrology techniques to achieve the HV performance (see Figure~\ref{fig:electrode}). Significant effort has been given to developing a sophisticated design of the support collars and standoffs for holding the deflecting plates stable whilst permitting micron-level adjustment of roll, yaw and radial position. 
A TiN coating less than 2 microns thick is being applied to the plates in collaboration with UK industry. It will ensure that the plates can operate at a sustained minimum level of \SI{4.4}{MV/m} without high-voltage breakdown and with significantly reduced dark current.

For the ultra-pure magnetic quadrupole, air-core quadrupoles are being assembled with the goal of keeping local dipole field errors below $10 \, \mu$T, corresponding to a $10 \, \mu$m mechanical tolerance, using electrical dipole correction via independently driven coils. A prototype of the assembly is shown in
Figure~\ref{fig:Mag_quad}, which is 40\,cm in length with an aperture 18\,cm (the coils are located outside the vacuum chamber). As can be seen, the quadrupole coils have been wound and, once assembly is complete, field measurements and correction tests will use beam position monitor data and a machine-learning-based linear regression model to estimate misalignment and adjust coil currents in real time. Its air-core design ensures high-fidelity, high-purity quadrupole field reversal at every beam injection to provide superior control of systematic uncertainties.

\section{Commitment to open communication}

The pEDM collaboration is committed to maintaining open, transparent, and regular communication with the broader scientific community as it advances its design, development, and implementation of a dedicated electric storage ring to measure the proton EDM with a sensitivity better than $10^{-29}\, e\, \cdot$cm. Given the complexity and interdisciplinary nature of this effort (spanning high-voltage technology, beam and spin dynamics, particle tracking, and systematic error control) clear and timely dissemination of both theoretical insights and experimental or conceptual progress is critical. To ensure this, the collaboration will establish annual in-person collaboration meetings, weekly general meetings, an open-access publication and preprint strategy, and community outreach and workshops. All communications will be publicly disseminated wherever possible. 

This proactive communication strategy is also designed to strengthen international collaboration, avoid duplication of effort, and accelerate progress toward a definitive measurement of the proton EDM. The collaboration recognizes that this measurement represents not only a technical challenge, but also a profound scientific opportunity, one that calls for openness, and sustained dialogue throughout the global community. 

\section{Summary}

The pEDM Experiment is a high-precision initiative aimed at directly measuring the EDM of the proton, with an initial sensitivity target of $10^{-29}\, e \cdot \text{cm}$, surpassing current limits by four orders of magnitude. It is poised to deliver a precision probe with the potential to uncover new sources of CP violation and probe physics beyond the Standard Model, including axion dark matter and the strong CP problem. With mature technologies, minimized systematics, and vibrant global collaboration, the experiment is technically ready to deliver a rich physics program that extends beyond the proton EDM alone (and can also measure the deuteron and $^3$He EDMs). It is complementary and mutually reinforcing to European efforts by the JEDI collaboration to develop prototype and full-scale deuteron/proton storage rings~\cite{CPEDM:2019nwp}, and the upcoming measurement of the muon EDM at PSI~\cite{Adelmann:2021udj}. Put simply, it will become a cornerstone of the global effort to explore fundamental symmetry violations and deepen our understanding of the universe. 

The experiment leverages a novel hybrid-symmetric, frozen-spin storage ring design using radial electric bending and magnetic focusing to suppress systematic errors. It allows for simultaneous clockwise and counterclockwise beam storage, enhancing cancellation of dominant systematic effects. The proposed ring, approximately 800 meters in circumference, would be installed at BNL, reusing the AGS tunnel to minimize infrastructure costs. Phase-I construction is projected to take 3--5 years, with data collection extending to 5 years for full statistical reach (without stochastic cooling) and a tenth of that with SC implemented. The collaboration, comprising institutions across the U.S., Europe, and Asia, is working toward a comprehensive CDR by 2026. Strong emphasis is placed on democratic collaboration, transparent communication, and community engagement.

pEDM stands as a vital, timely complement to large-scale collider programs and represents a compelling opportunity for early-career researchers.  It is a strategically complementary effort to CERN’s energy-frontier program, aligning with the goals of the Physics Beyond Colliders (PBC) initiative by offering a cost-effective path to probe new physics using existing infrastructure. As a timely, high-impact project with a relatively short timescale, pEDM can be a vital contribution to the 2026 European Strategy for Particle Physics Update, supporting continuity in research, training, and innovation, particularly for early-career researchers during gaps in major collider operations.

\microtypesetup{protrusion=false}
\sloppy
\printbibliography

\end{document}